\documentclass[letterpaper, 9 pt, conference]{ieeeconf} 

\IEEEoverridecommandlockouts                              

\usepackage{textcomp, libertine} 
\usepackage{amsmath,epsfig}
\usepackage{pgfplots}
\usepackage{amssymb}
\usepackage{amsxtra}
\usepackage{here}
\usepackage{cite}
\usepackage{dsfont}
\usepackage{graphicx}
\usepackage{epstopdf}
\usepackage{lettrine}
\usepackage{caption}
\usepackage{enumerate}
\usepackage[T1]{fontenc}
\newtheorem{theorem}{\bf Theorem}

\newtheorem{proposition}{\bf Proposition}

\newtheorem{definition}{\bf Definition}

\title{\LARGE \bf Mean-Field Games for Distributed Caching in Ultra-Dense Small Cell Networks\vspace{-0.2cm} }

\author{\authorblockN{ Kenza Hamidouche$^{1,2}$, Walid Saad$^2$, M\'erouane Debbah$^{1,3}$, and H. Vincent Poor$^{4}$} \authorblockA{\small
$^{1}$ CentraleSup\'elec, Universit\'e Paris-Saclay, France, {\tt\footnotesize kenza.hamidouche@centralesupelec.fr}\\
$^2$ Wireless@VT, Bradley Department of Electrical and Computer Engineering, Virginia Tech, USA, {\tt\footnotesize walids@vt.edu}\\
$^{3}$ Mathematical and Algorithmic Sciences Lab, Huawei France R\&D, France, {\tt\footnotesize merouane.debbah@huawei.com}\\
 $^{4}$ Electrical Engineering Department, Princeton University, Princeton, USA, {\tt\footnotesize poor@princeton.edu} \vspace{-0.6cm}}%
   \thanks{This research was supported by the ERC Starting Grant 305123 MORE (Advanced Mathematical Tools for Complex Network Engineering), and in part by the U.S. National Science Foundation under Grants CNS-1513697, CNS-1460333, CNS-1460316, and CCF-1420575.}
 }
\begin{document}

\maketitle
\begin{abstract}
In this paper, the problem of distributed caching in dense wireless small cell networks (SCNs) is studied using mean field games (MFGs). In the considered SCN, small base stations (SBSs) are equipped with data storage units and cooperate to serve users' requests either from files cached in the storage or directly from the capacity-limited backhaul. The aim of the SBSs is to define a caching policy that reduces the load on the capacity-limited backhaul links. This cache control problem is formulated as a stochastic differential game (SDG). In this game, each SBS takes into consideration the storage state of the other SBSs to decide on the fraction of content it should cache. To solve this problem, the formulated SDG is reduced to an MFG by considering an ultra-dense network of SBSs in which the existence and uniqueness of the mean-field equilibrium is shown to be guaranteed. Simulation results show that this framework allows an efficient use of the available storage space at the SBSs while properly tracking the files' popularity. The results also show that, compared to a baseline model in which SBSs are not aware of the instantaneous system state, the proposed framework increases the number of served files from the SBSs by more than 69\%. 
\end{abstract}
\section{Introduction}
Meeting the stringent quality-of-service (QoS) requirements of emerging wireless services requires significant changes to modern-day cellular systems \cite{CISCO2014}. One promising such change is through the dense and viral deployment of small base stations (SBSs) that can provide an effective way to boost the capacity and coverage of wireless networks. However, to benefit from this SBS deployment, several technical challenges must be addressed, in terms of interference management, resource allocation, and more importantly, backhaul management~\cite{andrews2014will}. 

Indeed, short range and low-power SBSs must be connected to the core network through backhaul links that are of limited capacity and can be owned by a third party \cite{andrews2014will}. Such capacity-limited and heterogeneous backhaul links through which the SBSs download the content can lead to significant delays when serving a large number of requests. One of the proposed solutions to cope with the backhaul bottleneck is via the use of \emph{distributed caching} at the network edge \cite{bacstuug2014living}. The idea of distributed caching is based on the premise of fitting the SBSs with storage devices while also exploiting the available storage at the user equipments (UEs) to reduce the load on the backhaul links. In particular, the SBSs can predict the users' requests for popular content and, then, download this content ahead of time in order to serve users locally, without using the backhaul. 

One of the main challenges in distributed caching is to define when and which files need to be cached at each SBS while minimizing the load on the backhaul links. In this regard, different solutions have been proposed in the literature \cite{bacstuug2014living,elbamby2014content,blaszczyszyn2014optimal,poularakis2014multicast,poularakis2013exploiting,poularakis2013approximation,liu2014cache,zhou2015greendelivery}. Two cases can be distinguished, encoded file caching and complete file caching. In complete file caching, the SBSs can only cache complete files, while in encoded file caching, the SBSs might store fractions of the files and cooperate to serve users. A number of works have focused on the complete file caching case such as in \cite{elbamby2014content,blaszczyszyn2014optimal,poularakis2014multicast }. In \cite{elbamby2014content}, an optimization problem is formulated in order to minimize the service delay of the UEs. The work in \cite{blaszczyszyn2014optimal} proposed a geographical caching approach aimed at maximizing the probability of finding the requested files at the SBSs. In \cite{poularakis2014multicast}, a multicast-aware caching approach is proposed to maximize the number of served requests for the same files via a single multicast transmission. On the other hand, the authors in \cite{poularakis2013exploiting} proposed a proactive encoded caching policy based on the mobility information of the users. The proposed caching policy aims to minimize the probability of serving fractions of the requested content from the core network. Similarly, in \cite{poularakis2013approximation}, a joint encoded caching and routing problem is formulated and then reduced to a tractable facility location problem. In \cite{liu2014cache}, the joint problem of power and encoded cache control is formulated using an optimization approach that aims to create more opportunities for serving users by cooperation between the SBSs that are equipped with multiple antennas.

Despite being interesting, none of these works consider a practical \emph{ultra dense} small cell network (SCN) which is expected to lie at the heart of emerging 5G cellular systems in which thousands of SBSs will be deployed within small geographical areas ~\cite{andrews2014will}. In fact, network density makes it difficult for the SBSs to coordinate and cache the files according to the state of all the other SBSs, which is necessary to prevent caching the same segments. Moreover, none of the existing works accounts for the realistic time-varying dynamics of the storage spaces.

The main contribution of this paper is to develop a novel approach to analyze the use of encoded caching in a network with a large number of SBSs. For a given SCN, we assume that each SBS has a state vector which is composed of the state of the wireless channel between the SBS and the served UE as well as the state of the storage unit. The dynamics of such a state vector is modeled via an It\^o process. Then, we formulate the cache decision problem as a stochastic differential game (SDG) in which the SBSs' goal is to maximize the number of requests served from the cache while taking into account the state of all the other SBSs. We show that, by considering a dense network of homogeneous SBSs, the SDG can be reduced to a mean-field game (MFG) \cite{lasry2006jeux,bensoussan2013mean,tembine2014risk}, in which the existence and uniquiness of the mean-field equilibrium are guaranteed. In this MFG, the individual state of the SBSs can be replaced by an average overall state, called the \emph{mean-field} which captures the global system state. Moreover, it is shown that deriving the equilibrium of the formulated MFG amounts to solving a coupled system of Hamilton-Jacobi-Bellman (HJB) and Fokker-Planck-Kolmogorov (FPK) equations. Simulation results show that the proposed framework allows an effective use of the storage space by modeling the dynamics of the storage spaces based on files' popularities. Moreover, in the proposed framework, the number of served requests from the SBSs can be increased by more than 69\% compared to a baseline, static model in which the SBSs are not aware of the instantaneous system state. 

The rest of the paper is organized as follows. Section \ref{sec:network} presents the system model. In Section \ref{sec:problem}, we formulate and analyze the stochastic differential game and the mean field game. Simulation results are presented in Section \ref{sec:results}. Finally, conclusions are drawn in Section \ref{sec:conclusion}. \vspace{-0.3cm}
\section{System Model}
\label{sec:network} 
Consider a wireless SCN composed of a set $\mathcal{N}$ of $N$ SBSs. The SBSs are equipped with data storage units that allow them to serve users' requests via radio links. Users can request videos from a set $\mathcal{V}$ of $V$ videos. When the requested files are not available in the storage units, users are served from the core network via backhaul links. The aim from equipping the SBSs with storage units is to reduce the traffic load of the capacity-limited backhaul links especially during peak hours. To this end, users' requests must be predicted before being requested and cached at the network edge to serve users locally via the neighboring SBSs, without using backhaul links. An illustration of the system model is given in Fig. \ref{fig:model}.

We consider a caching model similar to the one proposed in \cite{liu2014cache}, in which each video is encoded using an ideal maximum distance separable (MDS) rateless code, and then cached in the storage units. An MDS code divides the original file $k$ into segments of $q_k$ bits, and each segment is encoded into a longer sequence of parity bits, such that any arbitrary $q_k$ bits are sufficient to regenerate the original segment. 
This caching structure allows each node to control the number of bits to be stored in its storage. Here, a user can be served by more than one node at the same time without having to cache the whole file at all the serving nodes. Moreover, cooperation between SBSs may allow users to receive the requested content within a shorter time duration. The main challenge for each SBS $i$ is to define the fraction of each file $k$ that should be stored while optimizing a given cost. This is done by defining the values of the cache control variables $\boldsymbol{n}^{(i)}_{t}=[n^{(i)}_{1,t},...,n^{(i)}_{k,t},...,n^{(i)}_{V,t}]$, where $n^{(i)}_{k,t}\in [0,1], \forall k\in\mathcal{V}$ is the fraction of file $k$ that will be downloaded by node $i$ at time $t$. 

The wireless network is assumed to operate using a single frequency band, over which the two sets of SBSs and UEs request and exchange video files from the set $\mathcal{V}$ of videos. We assume that the users' requests follow a Zipf distribution which is a common model used for describing realistic file popularity distributions \cite{bacstuug2014living}. Thus, the probability of receiving a request for the $k^{\text{th}}$ video at the SBSs is given by the following probability mass function: 
\begin{equation*}
\Omega_t(k,V,\beta)=\frac{\frac{1}{k^{\beta}}}{\sum_{i=1}^{V}{\frac{1}{V}}},
\end{equation*}
where the parameter $\beta$ characterizes the steepness of the distribution. 
Based on this distribution, the nodes need to define a caching policy that allows them to download parts of the requested videos, while taking into account the network conditions such as the channel model as well as the internal state of each node described by the storage capacity \cite{alpcan2013mechanisms}. The instantaneous dynamics of these parameters are given next. 
\subsection{Channel model}
SBSs are configured in order to enhance the coverage of UEs by deploying them closer to the UEs. They can be deployed either by the operators in densely built-up areas or by the UEs inside the buildings. To capture the effects of the channel fading on the transmitted signal in such environments, we use the model in \cite{charalambous1999stochastic}, which showed that the instantaneous dynamics of non-stationary channel models can be described by a mean-reverting square process or a mean-reverting Ornstein-Uhlenbeck process. Here, we consider a time-varying additive Gaussian channel model which is given by $|h_{i,t}|^2$. We model the dynamics of the channel by a mean-reverting Ornstein-Uhlenbeck process which is a general model that describes additive Gaussian channels \cite{charalambous1999stochastic}. The dynamics of the channel are given by
\begin{equation}
~\textrm{d}h_{i,t}= \frac{\alpha}{2} (\mu_h- h_{i,t}) ~\textrm{d}t+\frac{1}{2}\sigma_h ~\textrm{d}\mathcal{B}_{i,t},
\end{equation}
where $\mu_h >0$, $\sigma_h>0$  and $\mathcal{B}_{i,t}$ is a standard Brownian motion.
\subsection{SINR model}
At a given time $t$, the signal received by a user from its serving SBS will experience interference from other SBS links. 
The signal-to-interference-plus-noise-ratio (SINR) is given by
\vspace{-0.1cm}

\begin{equation} 
\gamma_{i,t}=\frac{p_i|h_{i,t}|^2}{N_{0}+\frac{1}{N}\sum_{k\neq i, k\in \mathcal{N}}{p_k|h_{k,t}|^2}},
\end{equation}
where $p_i$ is the transmit power of SBS $i$, $N_{0}$ is the background noise level at the served user, and $\mathcal{N}$ is the set of all the potential transmitters. Analogous to \cite{de2012concurrent} and \cite{meriaux2013stochastic}, the interference from all the other SBSs is normalized to keep it bounded in a dense area of SBSs.
\subsection{Storage Unit Dynamics}
The available storage capacity at the SBSs changes stochastically depending on external control. Indeed, the storage of SBSs can be updated by either the users or the operator. 
Each UE is served by an SBS $i$ at an instantaneous rate given by
\begin{equation*}
 \kappa_{i,t}= \text{log}(1+\gamma_{i,t}).
\end{equation*}
We let $s^{(i)}_{k,t}$ be the number of bits from video $k$ that are stored at SBS $i$ at time $t$. To capture the randomness of users' behavior, we model the dynamics of the storage unit $s^{(i)}_{k,t}$ while serving file $k$ which was requested at $\boldsymbol{l}^{(i)}_{k}(x)$, as follows: 
\begin{align} 
~\textrm{d}s^{(i)}_{k,t}= & [n^{(i)}_{k,t} q_k- \beta (1-p_{k,t}) \bar{\zeta}_{i,t}]~\textrm{d}t+ \sigma_{s} ~\textrm{d}\mathcal{B}_{i,t},
\end{align}
where $n_{k,t}$ is the download rate of video file $k$ by the SBS and $q_k$ is the size of the file. The second term is the removal rate of file $k$ at the SBS, where $\beta>0$ is a parameter, $p_{k,t}$ is the popularity of file $k$ at time $t$, and $\bar{\zeta_{t}}$ is the mean number of bits downloaded by all the users up to time $t$. In fact, this term models the tradeoff between the popularity of the file and the number of bits that have been downloaded by the users in the time duration $[0,t]$. Here, we assume that there is no broadcasting and only one user can download a content item $k$ at time $t$ from a node $i$. 
$\zeta_{i,t}$ is the number of bits of file $k$ downloaded from node $i$ by the currently served user and is given by
\vspace{-0.4cm}

\begin{align}
\zeta_{i,t}= \text{min}{\bigg\{\kappa_{i,t}, s^{(i)}_{k,t_{l}}-\int_{t_{l}}^{t}{\zeta_{i,z}}~\textrm{d}z}+\int_{t_{l}}^{t}{n^{(i)}_{k,z} q_k}~\textrm{d}z \bigg\},
\end{align}
where $t_l=\boldsymbol{l}^{(i)}_{k}(x)$ is the time at which the SBS $i$ starts serving the requested file $k$. The number of downloaded bits is the minimum between the transmission rate from SBS $i$ to the served UE and the available bits from file $k$ at node $i$ which have not been downloaded yet by the served UE.\vspace{-0.1cm}
\section{Problem Formulation}
\label{sec:problem}
\vspace{-0.1cm}
The goal of an SBS $i$ is to decide on the values of the cache control variables and find the fraction $n^{(i)}_{k,t}\in [0, 1]$, of video $k$ that should be downloaded at time $t$ for serving users. Due to the limited capacity of the backhaul links, SBS $i$ cannot download more than $B^{(i)}_{t}$ bits from the core network, resulting in the \emph{backhaul download} constraint $\sum_{k\in \mathcal{V}} {B^{(i)}_{k,t}\leq B^{(i)}_{t}}$, with $B^{(i)}_{k,t}$ the maximum allocated backhaul for downloading the fraction $n^{(i)}_{k,t}$ of file $k$. Each SBS $i$ in the system aims at finding the optimal control vector $\boldsymbol{n}^{*(i)}_{t}=[n^{*(i)}_{1,t},...,n^{*(i)}_{V,t}]$, among the set of all admissible storage allocations $\mathcal{A}_i$, that optimizes a cost function which is defined next. 
\vspace{-0.2cm}
\subsection{SBSs' Cost Function}
 The goal for each SBS $i$ is to determine the values of the cache control variables that maximize the amount of cached bits for file $k$ subject to its storage capacity $o_i$, in terms of bits. We denote the instantaneous download profile of file $k$ at all the SBSs by $\boldsymbol{n}_{k,t}=[n_{k,t}^{(1)},..., n_{k,t}^{(N)}]$. 
The global cost at a given SBS is affected by the following factors. 
\begin{itemize}
 \item \emph{The inter-SBS redundancy cost}: This represents the cost of caching parts of a file $k$ knowing that this file was already cached by other SBSs. This cost is determined by the function $c^{(i)}_{k,t}: \mathds{R}^N \to \mathds{R}, \boldsymbol{n}_{k,t} \mapsto  c^{(i)}_{k,t}(\boldsymbol{n}_{k,t})$, which models the dependence between the cache decisions of all the SBSs. Since a UE can be served by many SBSs at the same time, the cache states of the other SBSs need to be considered. This will prevent the SBS $i$ from caching the same content as the files already cached by the other SBSs. Since the cache state of an SBS is a function of its own cache decision (see (4)), then, the cache decision $n_{k,t}^{(i)}$ of SBS $i$ depends implicitly on the cache decisions $\boldsymbol{n}^{(-i)}_{k,t}=  [n^{(1)}_{k,t},...,n^{(i-1)}_{k,t},n^{(i+1)}_{k,t},..., n^{(N)}_{k,t}]$ chosen by all the other SBSs. The function $c^{(i)}_{k,t}$ will be defined later in (13) (see Section 3.C).
 \item \emph{The in-SBS redundancy cost}: This represents the cost of caching the same bits from the same file at a given SBS. To avoid this, we limit the maximum number of cached bits from file $k$ at an SBS $i$ to the size of the file $q_k$. This constraint can be modeled by $\nu_i\begin{bmatrix}s_{k,t}^{(i)}-q_k\end{bmatrix}$, where $\nu_i$ is a constant.
 \item \emph{The backhaul cost}: This represents the cost of downloading a fraction $n_{k,t}^{(i)}$ of file $k$  in order to be cached. This fraction of file $k$ is downloaded through the allocated backhaul $B^{(i)}_{k,t}$ for that file. This cost is given by the following function: $g_t^{(i)}(n_{k,t}^{(i)}): \mathds{R}^{+} \to \mathds{R}, n_{k,t}^{(i)}\mapsto g^{(i)}_{t}(n_{k,t}^{(i)})$:
\vspace{-0.3cm}

 \begin{equation}
  g_t^{(i)}(n_{k,t}^{(i)})= \left\{
 \begin{aligned}
-\text{log}(B^{(i)}_{k,t}-q_kn^{(i)}_{k,t})  &\text{ if } n^{(i)}_{k,t}< \frac{B^{(i)}_{k,t}}{q_k},\\ 
+\infty & \text{ if } n^{(i)}_{k,t}\geq \frac{B^{(i)}_{k,t}}{q_k}.
\label{system}
\end{aligned}
  \right.
\end{equation}
Note that, the proposed framework can accommodate any other form of the backhaul cost function.
 \item \emph{The storage cost}: This represents the cost of storage at the SBS and allows modeling the limited storage capacity of the SBS $i$ which should not exceed $o_i$ bits. It is given by $\omega_i \begin{bmatrix}\sum_{k=1}^{V}s^{(i)}_{k,t}-o_i\end{bmatrix}$, where $\omega_i$ is a constant. 
\end{itemize}
Thus, the global cost function can be written as:

 \begin{align}  
J_{k,t}^{(i)}(n^{(i)}_{k,t}, \boldsymbol{n}^{(-i)}_{k,t})= & c^{(i)}_{k,t}(\boldsymbol{n}_{k,t})+ g_t^{(i)}( n^{(i)}_{k,t})+\nu_i\begin{bmatrix}s_{k,t}^{(i)}-q_k\end{bmatrix}\nonumber \\&+ \omega_i \begin{bmatrix}\sum_{k=1}^{V}s^{(i)}_{k,t}-o_i\end{bmatrix}.
\end{align}

Next, we formulate the cache control problem as a stochastic differential game.
\subsection{Stochastic Differential Game Formulation}
Let $\mathcal{N}$ be the set of \emph{players}. The state of a player $i$ at time $t$ with respect to a given file $k$ is defined as $\boldsymbol{y}^{(i)}_{k,t}=( h_{i,t}, s^{(i)}_{k,t})$, $\forall i\in\mathcal{N}, k \in \mathcal{V}$. The stochastic differential caching game is defined by
$(\mathcal{N}, (\mathcal{Y}_i)_{i\in\mathcal{N}}, \mathcal{A}_{i\in\mathcal{N}},(\mathcal{J}_i)_{i\in\mathcal{N}})$
where 
\begin{itemize}
\item $\mathcal{N}$ is the set of SBSs;
\item $\mathcal{Y}_i$ is the set space of SBS $i$ and follows
\begin{equation}
  \left\{
\begin{aligned}
~\textrm{d}h_{i,t}= &\frac{\alpha}{2} (\mu_h- h_{i,t}) ~\textrm{d}t+\frac{1}{2}\sigma_h ~\textrm{d}\mathcal{B}_{i,t},   \\ 
~\textrm{d}s^{(i)}_{k,t}= & [n^{(i)}_{k,t} q_k- \beta (1-p_{k,t}) \bar{\zeta}_{i,t}]~\textrm{d}t+ \sigma_{s} ~\textrm{d}\mathcal{B}_{i,t}.
\label{system}
\end{aligned}
  \right.
\end{equation}

\item$\mathcal{A}_i$ is the set of admissible caching control policies for node $i$; and
\item $\mathcal{J}_k^{(i)}$ is the cost function of node $i$ defined as follows:
\vspace{-0.2cm}

\begin{equation*}
\mathcal{J}^{(i)}_k=
\begin{aligned}
&\mathds{E}\begin{bmatrix}\displaystyle\int_{0}^{T}{J_{k,t}^{(i)}(n_{k,t}^{(i)}, \boldsymbol{n}^{(-i)}_{k,t}) ~\textrm{d}t}+ \psi^{(i)} (\lambda_T)\end{bmatrix},&
\label{system}
\end{aligned}
\end{equation*}
\end{itemize}
where the function $\psi^{(i)} (\lambda_T): [0,1] \to \mathds{R}, \lambda_T \to \psi^{(i)} (\lambda_T)$ models the cost of having a fraction $\lambda_T$ of free storage space at the end of the period $[0,T]$. This function guarantees that the SBS's owner will not use all the storage space but will keep a fraction for specific functionalities such as system updates or for the user's own usage when the SBS is owned by a user. 

Assume that, at each time $ t \in [0, T]$, a player $i$ can observe the current state $\underline{\boldsymbol{y}}_{k,t}=(\boldsymbol{y}^{(1)}_{1,t},...,\boldsymbol{y}^{(i)}_{k,t},..., \boldsymbol{y}^{(N)}_{k,t})$ of the system with respect to file $k$. However, this player has no additional information about the strategy of the other players. In particular, it cannot predict the future actions of the other players. In this case, the solution of the game can be captured via the following equilibrium concept:

\begin{definition}
A control strategy $\boldsymbol{n}_{t}^*$ is said to be a \emph{feedback Nash equilibrium} of the SDG if and only if $\forall i \in \mathcal{N}, \forall k \in \mathcal{V}, ~~ n^{(i)*}_{k,t}$ is the solution of the control problem
\begin{equation}
v^{(i)}_{k,t}(\underline{\boldsymbol{y}}_{k,t})= 
\begin{aligned}
 & \underset{n^{(i)}_{k,t}}{\text{inf}}
 & & \mathcal{J}_k^{(i)}.
 \end{aligned}
\end{equation}
$v^{(i)}_{k,t}(\underline{\boldsymbol{y}}_{k,t})$ is called the value function.
\end{definition}

A condition for the existence of a feedback Nash equilibrium for the SDG is the existence of a solution to the following HJB equations for each SBS $i$ and file $k$ \cite{oksendal2003stochastic}:
\vspace{-0.5cm}

\begin{align}
&\partial_t v^{(i)}_{k,t}(\underline{\boldsymbol{y}}_{k})+\begin{bmatrix}n^{(i)}_{k,t} q_k- \beta (1-p_{k,t}) \bar{\zeta}_{i,t}\end{bmatrix}\partial_s v^{(i)}_{k,t}(\underline{\boldsymbol{y}}_{k})\nonumber\\
&+\frac{\alpha}{2} (\mu_h- h_{i,t})\partial_h v^{(i)}_{k,t}(\underline{\boldsymbol{y}}_{k})+\frac{\sigma^2_S}{2} \partial_{ss}^2v^{(i)}_{k,t}(\underline{\boldsymbol{y}}_{k})\\
&+\frac{\sigma^2_h}{2} \partial_{hh}^2v^{(i)}_{k,t}(\underline{\boldsymbol{y}}_{k})+J^{(i)}_{k,t}(n^{(i)}_{k,t}, \boldsymbol{n}^{(-i)}_{k,t})=0.\nonumber
\end{align}
A sufficient condition for the existence and uniqueness of a solution $v_{k,t}^{(i)}(\boldsymbol{y}_{k})$ to the corresponding HJB equation is the smoothness of the drift functions of the dynamic equations and the cost function, i.e. the functions belong to $\mathcal{C}^{\infty}$ \cite{oksendal2003stochastic}. However, even if the choice of the functions in our system guarantees the existence and uniqueness of the solution, solving the $V\times N$ coupled HJB equations can be complex in a dense network of SBSs. Moreover, it is very difficult for a given SBS to observe all the states of the other nodes in a large scale wireless network. Interestingly, the analysis of the system becomes tractable using a mean-field approximation, when the number of players is considered very large. This solution is mainly practical for the emerging SCNs that are expected to include millions of connected devices. Thus, we will study the asymptotic case in the following section.
\subsection{Mean-Field Game Formulation}
We are interested in solving a stochastic optimal control problem when the number of SBSs is large ($N\rightarrow\infty$). To study the convergence of the system into the mean field, we assume that the state and download control preserve the exchangeability property which is defined as follows.

\begin{definition}
The states $\boldsymbol{y}^{(1)}_{k,t}, \boldsymbol{y}^{(2)}_{k,t},...,\boldsymbol{y}^{(N)}_{k,t}$ are said to be \emph{exchangeable} under the strategy $n^{(i)}_{k,t}$ if they generate a joint law which is invariant by permuting the SBSs' indices, i.e.,
\begin{equation*}
 \mathcal{L}(\boldsymbol{y}^{(1)}_{k,t},...,\boldsymbol{y}^{(N)}_{k,t}|n^{(i)}_{k,t}) =\mathcal{L}(\boldsymbol{y}^{({\pi(1)})}_{k,t},...,\boldsymbol{y}^{({\pi(N)})}_{k,t}),
\end{equation*}
for any bijection $\pi$ defined over $\{1,...,N\}$.
\end{definition}

To guarantee this property, we make the following assumptions: 
\begin{itemize}
\item Each SBS knows its individual state; and
\item Each SBS implements a homogeneous caching policy: $n^{(i)}_{k,t}= f_k(t,\boldsymbol{y}^{(i)}_{k,t})$.
\end{itemize}
Due to this exchangeability property, all the players become indistinguishable and thus we can focus on a generic SBS whose state is now given by $\boldsymbol{y}_{k,t}=[h_t, s_{k,t}]$. Under the exchangeability property, we can simplify the previous system of coupled HJB equations by considering that a given player defines its control policy based only on its state and the mean-field. By considering such a system, a player does not require the knowledge of each player's state in the system but only the distribution of those players over the states. The convergence of the SDG to a mean field game is provided in he following result.
\begin{theorem}
Define $M^N_{k,t}=\frac{1}{N}\sum_{i=1}^{N}{\delta_{\boldsymbol{y}_{k,t}^{(i)}}}$ as the occupancy measure of the $N$ SBSs. Suppose that the states $\boldsymbol{y}^{(i)}_{k,t}$ and the caching control $n^{(i)}_{k,t}$ preserve the exchangeability property, then $M^N_{k,t}$  converges in distribution to $m_k$. Moreover, the law $m_{k,t}$ is the solution of the following FPK equation:
\begin{align*}
&m_{k,0}(\boldsymbol{y}_{k})=\rho_0(\boldsymbol{y}_{k}), \forall \boldsymbol{y}_{k}\\
&\partial_t m_{k,t}(\boldsymbol{y}_{k})+ \begin{bmatrix}n_{k,t} q_k- \beta (1-p_{k,t}) \bar{\zeta_{t}}\end{bmatrix}\partial_s m_{k,t}(\boldsymbol{y}_{k})\\
&+\frac{\alpha}{2}(\mu- h_{t})\partial_h m_{k,t}(\boldsymbol{y}_{k})-\frac{\sigma^2_s}{2} \partial_{ss}^2 m_{k,t}(\boldsymbol{y}_{k})\\
&-\frac{\sigma^2_h}{2} \partial_{hh}^2m_{k,t}(\boldsymbol{y}_{k})=0.
\end{align*}
\end{theorem}
\vspace{0.2cm}
\begin{proof}
The proof is given in Appendix $\rm I$.
\end{proof}

Under the exchangeability property and for $N \rightarrow\infty$ and $\forall k \in \mathcal{V}$ we have, $\gamma_{i,t}\rightarrow \gamma_t$, which can be derived as in \cite{meriaux2013stochastic}.

Hence, the dynamics of the state for a generic SBS can be defined by the following system of differential equations:
\begin{equation}
  \left\{
\begin{aligned}[left]
~\textrm{d}h_{t}=    & \frac{\alpha}{2} (\mu - h_{t}) ~\textrm{d}t+\sigma_h ~\textrm{d}\mathcal{B}_{t},\\
~\textrm{d}s_{k,t}=& \begin{bmatrix}n_{k,t} q_k- \beta (1-p_{k,t}) \bar{\zeta_{t}}\end{bmatrix}~\textrm{d}t+ \sigma_{s} ~\textrm{d}\mathcal{B}_{t}. 
\label{system}
\end{aligned}
  \right.
\end{equation}
The aim for each node is to choose a caching control $ n_{k,t}$ for each file $k\in \mathcal{V}$  in order to minimize the following cost function:
\begin{equation}
\mathcal{J}_k=
\begin{aligned}
&\mathds{E}\begin{bmatrix}\displaystyle\int_{0}^{T}{J_{k,t}(n_{k,t}, m_{k,t}) ~\textrm{d}t} \end{bmatrix}+\psi(\lambda_T).& 
\label{system}
\end{aligned}
\end{equation}
Now we can redefine the cost function as a function of the mean field process. In this regard, we define the function $c_{k,t}(m_{k,t})$ which is now a function of the mean field process $m_{k,t}$. Intuitively, the larger is the expected cached fractions of file $k$ in the network, given by $\phi_k=\frac{\textrm{d}}{\textrm{d}t}\int_{0}^{q_k}{s_{k}m_{k,t}}(\boldsymbol{y}_{k})\textrm{d}h$, the lower is the interest of the SBS in caching bits from that file. However, when the mean number of cached bits of file $k$ is low, the cost should be defined in order to encourage the SBS to cache bits of file $k$ until a given threshold. On the other hand, an SBS would aim to cache more bits from file $k$ when the expected requests for that file is high. Thus, the cost depends as well on the distribution of users' requests. The cost can then be written as 

\vspace{-0.4cm}
\begin{equation} 
c_{k,t}( m_{k,t})= \exp(-\varrho_1 \phi_k)+\frac{\varrho_2 \phi_k}{\Omega_t(k,V,\beta)},
\end{equation}
where $\varrho_1$ and $\varrho_2$ are constants.

The value function for a generic SBS is given by
\begin{equation}
v_{k,t}(\boldsymbol{y}_k)=
  \begin{aligned}
 & \underset{n_{k,t}}{\text{inf}}
 & & \{\mathcal{J}_{k}\}.
 \end{aligned}
\end{equation}

Finding the optimal control of a given SBS and file $k$ amounts to jointly solving the following mean field problem:

\begin{align*}
&\partial_t v_{k,t}(\boldsymbol{y}_{k})+\begin{bmatrix}n_{k,t} q_k- \beta (1-p_{k,t}) \bar{\zeta_{t}}\end{bmatrix}\partial_s v_{k,t}(\boldsymbol{y}_{k})\\
&+\frac{\alpha}{2} (\mu_h- h_{t})\partial_h v_{k,t}(\boldsymbol{y}_{k})+\frac{\sigma^2_s}{2} \partial_{ss}^2v_{k,t}(\boldsymbol{y}_{k})+\frac{\sigma^2_h}{2} \partial_{hh}^2v_{k,t}(\boldsymbol{y}_{k})\\
&+J_{k,t}(n_{k,t},m_{k,t})=0,
\end{align*}
\begin{align*}
&m_{k,0}(\boldsymbol{y}_{k})=\rho_0(\boldsymbol{y}_{k}), \forall \boldsymbol{y}_{k},\\
&\partial_t m_{k,t}(\boldsymbol{y}_{k})+ \begin{bmatrix}n_{k,t} q_k~- \beta (1-p_{k,t}) \bar{\zeta_{t}}\end{bmatrix}\partial_s m_{k,t}(\boldsymbol{y}_{k})\\
&+\frac{\alpha}{2}(\mu -h_{t})\partial_h m_{k,t}(\boldsymbol{y}_k)-\frac{\sigma^2_s}{2} \partial_{ss}^2m_{t}(\boldsymbol{y})\\
&-\frac{\sigma^2_h}{2} \partial_{hh}^2m_{k,t}(\boldsymbol{y}_{k})=0.
\end{align*}

The advantage of the mean field formulation is that SBSs do not need full knowledge of the state or the caching strategy of other SBSs to compute the outcome of the game. Also, in order to find the optimal power control an SBS has to solve only one HJB equation for a given file. 
To compute the optimal download control, we have to solve the two coupled equations in $v_k$ and $m_k$. From the optimization standpoint, finding the solution of the stochastic optimal control amounts to finding the optimal caching control that minimizes the Hamiltonian. The Hamiltonian function is defined as follows:
 \begin{align}
& H(\boldsymbol{y}_k,m_{k,t},\nabla v_k)=\{\frac{\alpha}{2} (\mu_h- h_{t})\partial_h v_{k,t}(\boldsymbol{y}_{k})\nonumber \\
&+\begin{bmatrix}n_{k,t} q_k- \beta (1-p_{k,t}) \bar{\zeta_{t}}\end{bmatrix}\partial_s v_{k,t}(\boldsymbol{y}_{k})\}+J_{k,t}(n_{k,t},m_{k,t})
 \end{align}
with $\nabla v_k$ the gradient. 

The optimal value of the number of bits that should be downloaded by each SBS is given in the following result. 
\begin{proposition}
The optimal control $n^*_{k,t}$ of the stochastic mean field game is given by
\begin{align} 
n^*_{k,t}=\frac{1}{q_k}\begin{bmatrix}B_{t,k}-\frac{1}{2 \partial_s v_{k,t}(\boldsymbol{y}_{k})}\end{bmatrix}.
\end{align}
\end{proposition}
\vspace{0.2cm}
\begin{proof}
The proof is given in Appendix $\rm II$.
\end{proof}
\section{Numerical Results}
\label{sec:results}
\begin{figure}
\centering
\includegraphics[scale=0.44]{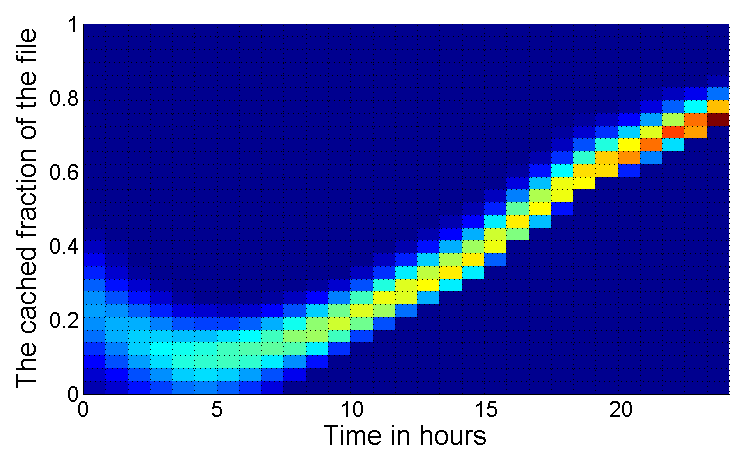}\vspace{-0.1cm}
\caption{Density solution $m^*(s_k,t)$ as a function of time $t$ and the cached fraction of file $k$.}\vspace{-0.6cm}
\label{fig:density}
\end{figure}

To solve the HJB-FPK system of equations, we proceed by solving iteratively the two equations using a simple fixed-point algorithm until convergence. We assume a static channel model and thus the state is only defined by the cache state of the SBSs. The transmit power is set to $p=1$ W and the noise to $N_0=-80$ dBm. In Fig. \ref{fig:density}, we show the evolution of $m^*$ for one file whose popularity increases over $24$ hours. For this, the file size is normalized to 1 and the storage capacity of the SBSs is set to $o=2/5$. The initial distribution of the SBSs follows a normal distribution $\mathcal{N}(0.2,0.1)$. Fig. \ref{fig:density} shows that the cached fraction of the file at the SBSs decreases when the file's popularity is low, thus making the storage space available for other more popular files. When more requests are expected for that file, all the SBSs cache a higher fraction of the file, in the limit of their storage capacity. This allows all the users to be served by any random subset of SBSs in their proximity, improving the experienced quality in terms of download time.

\begin{figure}
\centering
\includegraphics[scale=0.42]{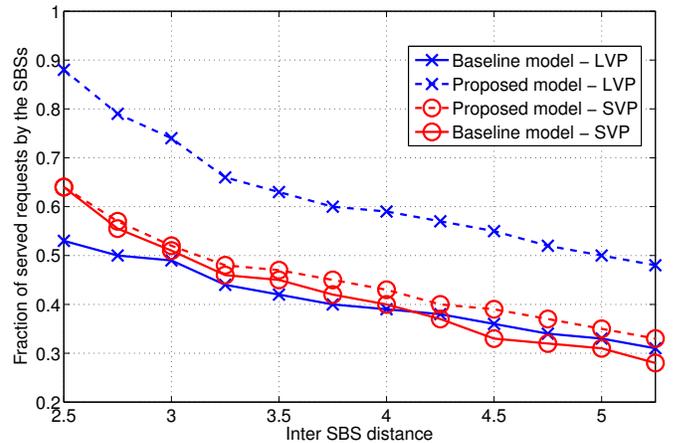}\vspace{-0.1cm}
\caption{Fraction of served requests from the BSBs with respect to the variation of popularity over time and inter SBS distance}\vspace{-0.6cm}
\label{fig:perf}
\end{figure}

In Fig. \ref{fig:perf}, we compare the proposed framework with a baseline method in which the SBSs are not aware of the instantaneous global state of the system. To define their optimal download rate and minimize the cost, the SBSs make their decisions based on the information from their previous experience by averaging the storage state over the past time periods. The two methods are compared in terms of the fraction of satisfied requests from the cache of the SBSs without using the backhaul with respect to the inter SBS site distance that models the density of the SBSs in the network. That is, the smaller is the distance between the SBSs, the denser is the network; here, one unit of inter SBS distance is set equal to 25 m. The comparison is performed for two different cases. The first case corresponds to scenarios that exhibit a large variability in the file popularity (LVP), from a given time period to another. In the second case, the popularity is assumed to change slightly (SVP). We can see from Fig. \ref{fig:perf}, that the denser the network, the higher is the number of requests served by the SBSs. In fact ,densifying the networks results in a given user being in the coverage area of a large number of SBSs which increases the probability of finding the requested file at one of the SBSs in its vicinity. The performance of the baseline model is closer to the proposed model when the popularity of the files changes slightly. This is due to the fact that the SBSs in this model update their storage space based on the past information which remains valid when the popularity changes slightly. However, the proposed algorithm outperforms the baseline model by increasing the number of served requests from the SBSs by up to 69\%, when the popularity of the files varies largely. In this case, fewer requests are served from the SBSs in the baseline model due to the significant changes in the SBSs' states over time, which does not allow the SBSs to adapt their control variables according to the real system state.
\vspace{-0.3cm}
\section{Conclusions}
\label{sec:conclusion}
\vspace{-0.1cm}
In this paper, we have studied the problem of distributed caching in ultra-dense small cell networks. We have formulated the cache control problem as a mean-field game in which the SBSs aim to minimize a given cost function while taking into account the cache state of all the other SBSs in the network. We have analyzed and showed the existence and uniqueness of the mean field equilibrium which is the solution of a coupled system of HJB and FPK equations. We have showed through simulations, that the proposed method enables efficient use of the storage space efficiently by adapting the control variable to the popularity of the files and the global state of the storage spaces in the network. Moreover, the proposed approach significantly decreases the load on the backhaul links by serving more requests locally from the SBSs.    
\bibliographystyle{IEEEtran}
\vspace{-0.5cm}
\bibliography{references}
\vspace{-0.4cm}
\appendices
\section{Proof of Theorem 1}
To prove the weak convergence of the occupancy measure $M_t^N$ to the mean field process $m$ in the caching control problem, we use Theorem 25.10 in \cite{billingsley2008probability} which gives a necessary and sufficient condition for weak convergence. 
Theorem 25.10 which is a result of Helly's Theorem, states that tightness is a necessary and sufficient condition for weak convergence. Thus, since $(\mathds{R}^2, \mathcal{B}(\mathds{R}^2) )$ is a separable complete metric space, then every sequence of probability measure defined on this space is tight, which leads to the weak convergence of $M_t^N$ to the mean field process $m$.
Next we derive the FPK equation which describes the evolution of the density of users per state. In what follows, we omit the file index $k$ for ease of notation and the proof is applicable $\forall k \in \mathcal{V}$: 
\begin{align*}
&m_{0}(\boldsymbol{y})=\rho_0(\boldsymbol{y}), \forall \boldsymbol{y}\\
&\partial_t m_{t}(\boldsymbol{y})+ \hat{u}_1 (\boldsymbol{y},t) \partial_s m_{t}(\boldsymbol{y})+ \hat{u}_2 (\boldsymbol{y},t) m_{t}(\boldsymbol{y})\\
&+\hat{\sigma}_1 \partial_{ss}^2 m_{t}(\boldsymbol{y})+\hat{\sigma}_2 \partial_{hh}^2m_{t}(\boldsymbol{y})=0,
\end{align*}
where $\hat{u}_1 (\boldsymbol{y},t)=- \begin{bmatrix}n_{k,t} q_k- \beta (1-p_{k,t}) \bar{\zeta_{t}}\end{bmatrix}$, $\hat{u}_2 (\boldsymbol{y},t)=\frac{\alpha}{2}(\mu- h_{t})$, $\hat{\sigma}_2=\frac{\sigma^2_h}{2}$ and $\hat{\sigma}_1=\frac{\sigma^2_s}{2}$. Let $\varphi(\boldsymbol{y},t)$ be a test function that belongs to $\mathcal{C}^{2}$ ($\mathds{R}^2)$ in space and $\mathcal{C}^{1}(\mathds{R})$ in time. By applying Ito's lemma we obtain
\begin{align}
\textrm{d}\varphi(\boldsymbol{y},t)  = &\sum_{i=1}^{2}{\hat{f}_i\frac{\partial \varphi}{\partial \boldsymbol{y}_i }}(\boldsymbol{y},t) +\sum_{i,j=1}^{2}{\hat{\sigma}_{ij}\frac{\partial \varphi}{\partial \boldsymbol{y}_i \partial  \boldsymbol{y}_j}(\boldsymbol{y},t) }\nonumber\\
&+ \sum_{i=1}^{2}{\sigma_i\frac{\partial \varphi}{\partial \boldsymbol{y}_i }(\boldsymbol{y},t)}\textrm{d}\mathcal{B}_i,
\end{align}
where $\hat{f}=(\hat{u}_1 (\boldsymbol{y},t), \hat{u}_2 (\boldsymbol{y},t))$ and $\hat{f}_i$ represents the $i$th entry of the vector $\hat{f}$. $\hat{\sigma}= \frac{1}{2}(\sigma_1,\sigma_2) (\sigma_1,\sigma_2)^T$ and $\hat{\sigma}_{ij}$ represents an entry of the matrix $\hat{\sigma}$.
By taking the expectation value on both sides we obtain
\begin{align}
\mathds{E}\textrm{d}\varphi(\boldsymbol{y},t)  = &\mathds{E}\sum_{i=1}^{2}{\hat{f}_i\frac{\partial \varphi}{\partial \boldsymbol{y}_i }}(\boldsymbol{y},t) +\mathds{E}\sum_{i,j=1}^{2}{\hat{\sigma}_{ij}\frac{\partial \varphi}{\partial \boldsymbol{y}_i \partial  \boldsymbol{y}_j}(\boldsymbol{y},t) }.
\end{align}

Because Brownian motion is a martingale, thus
$\mathds{E} \sum_{i=1}^{2}{\sigma_i\frac{\partial \varphi}{\partial \boldsymbol{y}_i }(\boldsymbol{y},t)}\textrm{d}\mathcal{B}_i=0$. By definition we have 
$\mathds{E}[\varphi(\boldsymbol{y},t)] = \int_{\mathds{R}^2} \varphi(\boldsymbol{x},t) m_{k,t}(x) \textrm{d}x$.
By substituting in (17), we obtain
\begin{align*}
 \int_{\mathds{R}^2}{  \varphi (\boldsymbol{y},t) m_{t}(\boldsymbol{y}) d y}=& \int_{\mathds{R}^2}{ \bigg [  \sum_{i=1}^{2}{\hat{f}_i\frac{\partial \varphi}{\partial \boldsymbol{y}_i }}(\boldsymbol{y},t)}\nonumber\\
&+\sum_{i,j=1}^{2}{\hat{\sigma}_{ij}\frac{\partial \varphi}{\partial \boldsymbol{y}_i \partial  \boldsymbol{y}_j}(\boldsymbol{y},t) }\bigg] m_t(\boldsymbol{y}) dy.
\end{align*}

After integrating by parts on the right-hand side we get

\begin{align}
\int_{\mathds{R}^2}{  \varphi (\boldsymbol{y},t) m_{t}(\boldsymbol{y}) d y}=& \int_{\mathds{R}^2}{ \bigg [ \sum_{i=1}^{2}{ \hat{f}_i \frac{\partial m_t}{\partial \boldsymbol{y}_i} (\boldsymbol{y})\varphi (\boldsymbol{y}_t)}}\nonumber\\
&- \sum_{i,j=1}^{2}{ \hat{\sigma}_{ij} \frac{\partial  m_t}{\partial{\boldsymbol{y}_i} }(\boldsymbol{y}) \frac{\partial \varphi}{\partial{\boldsymbol{y}_j}}(\boldsymbol{y},t)}\bigg]  dy.
\end{align}

A further integration by part of the second term on the right-hand side gives
\begin{align}
\int_{\mathds{R}^2}{  \sum_{i,j=1}^{2}{ \hat{\sigma}_{ij}\frac{\partial  m_t }{\partial_{\boldsymbol{y}_i} }(\boldsymbol{y}) \frac{\partial \varphi}{\partial_{\boldsymbol{y}_j}}(\boldsymbol{y},t)}}= \nonumber\\
- \int_{\mathds{R}^2}\sum_{i,j=1}^{2}{ \hat{\sigma}_{ij}\frac{\partial^2 m_t}{\partial \boldsymbol{y}_i \partial \boldsymbol{y}_j}} (\boldsymbol{y})  \varphi (\boldsymbol{y},t) dy.
\end{align}

By substituting in (18), we have
\begin{align}
\int_{\mathds{R}^2}{\varphi (\boldsymbol{y}) \bigg[  m_{t}(\boldsymbol{y})-\sum_{i=1}^{2}{ \hat{f}_i \frac{\partial m_t}{\partial \boldsymbol{y}_i}(\boldsymbol{y})} }\nonumber\\
-\sum_{i,j=1}^{2}{\hat{\sigma}_{ij}\frac{\partial^2 m_t}{\partial \boldsymbol{y}_i \partial \boldsymbol{y}_j}} (\boldsymbol{y})  \bigg ] dy=0.
\end{align}
Then using the generalized variational lemma we have that [lemma 7.1.2] \cite{atkinson2009theoretical}
\begin{align}
&m_{t}(\boldsymbol{y})-\sum_{i=1}^{2}{ \hat{f}_i \frac{\partial m_t}{\partial \boldsymbol{y}_i}(\boldsymbol{y})} -\sum_{i,j=1}^{2}{\hat{\sigma}_{ij}\frac{\partial^2 m_t}{\partial \boldsymbol{y}_i \partial \boldsymbol{y}_j}} (\boldsymbol{y})=0,
\end{align}
which completes the proof.
%
%

\section{Proof of proposition 1}
\vspace{-0.5cm}
 \begin{align}
& H(\boldsymbol{y}_k,m_{k,t},\nabla v_k)=\{\begin{bmatrix}n_{k,t} q_k~- \beta (1-p_{k,t}) \bar{\zeta_{t}}\end{bmatrix}\partial_s v_{k,t}(\boldsymbol{y}_{k})\nonumber \\
&+\frac{\alpha}{2} (\mu_h- h_{t})\partial_h v_{k,t}(\boldsymbol{y}_{k})+J_{k,t}(n_{k,t},m_{k,t})\}.
 \end{align}
 The derivative of  $H(\boldsymbol{y}_k,m_{k,t},\nabla v_k)$ with respect to $n_{k,t}$ gives
\begin{align}
\partial_n  H(\boldsymbol{y}_k,m_{k,t},\nabla v_k)= q_k \begin{bmatrix}\partial_s v_{k,t}(\boldsymbol{y}_{k})+\frac{1}{B_{k,t}-q_k n_{k,t}}\end{bmatrix}.
\end{align}
By setting it to 0, we get $n_{k,t}^{*}$ which can be written as follows:
\begin{align}
n_{k,t}^{*}=\frac{1}{q_k}\begin{bmatrix} B_{k,t}+\frac{1}{\partial_s v_{k,t}(\boldsymbol{y}_{k})}\end{bmatrix}.
\end{align}

\end{document}